\documentclass[preprint,aps,superscriptaddress]{revtex4}
\newcommand{\memo}[1]{}
\usepackage{graphicx}
\usepackage{bm}
\usepackage{amsmath}
\usepackage{amssymb}
\usepackage{amsfonts}
\usepackage{fancybox}
\usepackage[dvips]{color}
\begin{document}

\preprint{KYUSHU-HET-120}

\title{Vacuum Structures of Supersymmetric Noncompact Gauge Theory}

\author{Kenzo Inoue}
\email{inoue@phys.kyushu-u.ac.jp}
\affiliation{Department of Physics, Kyushu University 
Fukuoka 812-8581 Japan}
\author{Hirofumi Kubo}
\email{kubo@higgs.phys.kyushu-u.ac.jp}
\affiliation{Synchrotron Light Application Center, Saga University 
1 Honjo, Saga 840-8502, Japan}
\author{Naoki Yamatsu}
\email{yamatsu@higgs.phys.kyushu-u.ac.jp}
\affiliation{Department of Physics, Kyushu University 
Fukuoka 812-8581 Japan}

\date{\today}

\begin{abstract}
We consider models with a noncompact symmetry in the framework of
$\mathcal{N}=1$ supersymmetry.  Contrary to the conventional approach,
the noncompact symmetry is realized linearly on all fields without
constraints.  The models are constructed using noncanonical K\"ahler
function and gauge kinetic function, which is introduced for the local
case. It is explained that the symmetry needs to be spontaneously broken
for the consistency of a model. We study the vacuum structures of two
models with the noncompact symmetry $SU(1,1)$ for both global and local
cases.  One of them includes two fundamental representations of the
group and the other includes one adjoint representation. It is shown
that the former is consistent for the global case and the latter is
consistent for both the global and local cases.
\end{abstract}
\maketitle

\section{Introduction}
\label{sec:intro}
Two of the present authors have studied a supersymmetric vectorlike
model based on $\mathcal{N}=1$ supersymmetry and a horizontal symmetry
$G_H$ \cite{Maehara:1979kf,Wilczek:1978xi,Froggatt:1978nt}, which
governs the generational structures of quarks and leptons
\cite{Inoue:1994qz,Inoue:1996te,Inoue:2000ia,Inoue:2003qi,Inoue:2007uj,Inoue:2008ng}.
It is constructed to reproduce the (constrained) minimal supersymmetric
standard model
(MSSM)~\cite{Inoue:1982pi,Inoue:1983pp,Ibanez:1982fr,Martin:1997ns,Chung:2003fi}
at low energies. This model seems to be a very promising candidate for
giving the explanation of the physics beyond the standard model and the
MSSM.  It has several distinguishing features: The appearance of three
chiral generations of quarks and leptons observed in Nature is explained
as a result of a dynamical phenomenon, spontaneous breaking of the
noncompact horizontal symmetry $G_H=SU(1,1)$. It also naturally explains
the hierarchical structure of the Yukawa couplings as a consequence of
the symmetry property of the group.  Furthermore, the model gives rise
to the violation of $P$, $C$ and $T$ symmetries observed in experiments
also as a consequence of the spontaneous breaking of $SU(1,1)$ gauge
symmetry, while being the exact symmetries of the model.  In this
scenario, the noncompact gauge symmetry plays an extremely important
role.

Although the model has several intriguing features, there remain some
important elements that need clarification. One of them concerns the
issue of whether a gauge theory based on a noncompact group can be
constructed at all.  Most of the studies of gauge theories so far are
devoted to the ones with compact groups and many of the important
properties have been understood.  In contrast, it seems that only a
little is understood about theories with noncompact internal symmetry
groups.  Among them, it is known that theories based on a noncompact
group such as $SU(1,1)$, global or local, are potentially afflicted with
so-called the ``ghost'' problem \cite{Julia:1979pv}, which could lead to
the breakdown of such theories.

To explain the problem briefly, let us consider the gauge fields
associated with $SU(1,1)$, which belong to the adjoint representation of
the group.  The ``canonical'' kinetic term for the fields is given as
\begin{eqnarray}
{\cal L}_{\rm gauge~kin.}
&=&
\frac{1}{4}\eta^{(3)}_{AB}F_{\mu\nu}^AF^{\mu\nu B},
\label{gaugekin}
\end{eqnarray}
where the indices $A$ and $B$ run over the adjoint representation and
$\eta^{(3)}_{AB}$ is the metric of $SU(1,1)$ in the adjoint
representation(or Killing form) defined as
\begin{eqnarray}
\eta^{(3)}_{AB}
\equiv
\left(
\begin{array}{ccc}
-1&0&0\\
0&-1&0\\
0&0&+1\\
\end{array}
\right),
\end{eqnarray}
and $F^A_{\mu\nu}$ are the field-strength tensor. We see immediately
that the metric in the kinetic term is not positive-definite.  The third
component $(A=3)$ has a sign opposite to the first and second ones
$(A=1,2)$, which have the standard sign.  It should be emphasized that
the actual problem is the simultaneous occurrence of the different
signs.  This kind of kinetic term alone would give rise to the
perturbative quanta with negative norm, which we refer to as a ``ghost''
in this paper, and thus we ends up with the Fock space with an
indefinite metric
\footnote{Note that the ``ghost'' that we are concerned here is
essentially different from the Faddeev-Popov ghosts, which are introduced
in the gauge-fixing procedure and can be shown to be absent from the
physical Hilbert space.}
.  In such theories, unitarity is violated, and therefore we would have
difficulty in making probabilistic interpretations in quantum theory, at
least in perturbation theory. Another problem is that the Hamiltonians
may be unbounded from below. If a ghost appears in a theory, the energy
of the system may decrease arbitrarily as more ghosts are created, and
thus its appearance implies the absence of a ground state.  One might
notice that these two problems are not independent. If a ground state
does not exist, one can not define any other states.  Hence, it would
not make much sense to dispute about the violation of unitarity to begin
with.  The situations are basically the same for fields of other spin
and in other representations of the group.  Therefore, we would
encounter the same problem also in the case of a global symmetry.

In the case of a global symmetry, a prescription to construct a theory
based on a noncompact group, which is now a standard one, is worked out
and explained in Ref.~\cite{Cremmer:1979up}.  In this construction, a
noncompact global group $G$ is spontaneously broken to its maximal
compact local subgroup $H$ so that $G$ is realized nonlinearly while $H$
is realized linearly (hidden local symmetry \cite{Bando:1987br}).  The
scalar fields in the theory correspond to Nambu-Goldstone(NG) bosons
that parametrize the coset space $G/H$, i.e. nonlinear sigma model.  The
essential point here is that the non-propagating ``gauge'' fields
associated with local group $H$ exactly cancel the kinetic terms for the
scalar fields that correspond to the ghosts. In other words, the
apparent ghost degrees of freedom are not dynamical and simply absent
right from the beginning.  There are many studies along this line on the
nonlinear sigma models based on noncompact group including the
evaluation of quantum corrections
\cite{Amit:1983eg,Cohen:1983xv,Davis:1983fs,Davis:1983rn,
Ohta:1983yx,vanHolten:1983cn,vanHolten:1984ia,Davis:1985kz}.

In the case of a noncompact local group, the ghosts may appear in gauge
fields as well as in fields of other spin.  In Ref.~\cite{deWit:1983xe},
de Wit et al. study such case in the framework of $\mathcal{N}=2$
supergravity.  In order to avoid the appearance of ghosts, they propose
to use a nonlinear multiplet of the group and to use the field that is a
part of the multiplet as a compensator. An important point in their
construction is the use of the framework of $\mathcal{N}=2$ supergravity
that is reduced from the conformally invariant one by imposing certain
gauge conditions \cite{deWit:1984pk}.  The compensating field is
extended to a supermultiplet, which contains a field that corresponds to
a ghost.  After the elimination of the tensor auxiliary field, the sign
of the kinetic term for the gauge fields is reversed and the ghost
disappears from the theory.

In this paper, we explore the possibility of constructing a sensible
theory with a {\it linearly} realized noncompact {\it local} group in
the framework of $\mathcal{N}=1$ global supersymmetry.  The basic idea
resembles that of Ref.~\cite{deWit:1983xe}. However, there are essential
differences: We consider a theory with $\mathcal{N}=1$ supersymmetry. A
noncompact symmetry is realized linearly with all fields being
independent dynamical degrees of freedom, i.e. without imposing the
constraints that reduce the degrees of freedom, which are used to define
nonlinear sigma models. Our construction does not need to involve
auxiliary fields nor compensators, which play essential roles in the
model studied in Ref.~\cite{deWit:1983xe}
\footnote{The group $SU(1,1)$ considered in the present paper is the
covering group of $SO(2,1)$~\cite{Gourdin} considered in
Ref.~\cite{deWit:1983xe}.  There is no essential difference between the
two.}
. We show that a gauge theory based on a noncompact group can be defined
as a sensible theory by presenting an explicit model. In this model, the
symmetry is realized linearly on all the unconstrained fields. We also
study the structure of the vacuum that are far more richer than those in
nonlinear sigma models and its relation to the ghost problem.

We briefly sketch out the scenario for constructing a ghost free theory
with a noncompact group in both global and local cases.  To be specific,
the ghost problem arises in theories with nonunitary representations of
a noncompact group, e.g. finite-dimensional representations. In this
light, we focus our attentions on theories with nonunitary
representations.  The essential points in the scenario are the
introduction of field-dependent kinetic terms for the fields and the
necessity of the spontaneous breaking of the noncompact group. This
seems to be the only possibility for realizing a sensible theory of this
kind, which means that a simple free theory can not be defined in the
case of a noncompact group.

Suppose that there exists a scalar field $\phi$ that belongs to some
representations of the group.  The gauge invariant kinetic term for the
gauge fields is given as
\begin{eqnarray}
{\cal L}_{\rm gauge~kin.}
&=&
-\frac{1}{4}f_{AB}(\phi)F_{\mu\nu}^AF^{\mu\nu B},
\end{eqnarray}
where the indices $A$ and $B$ run over the adjoint representation of the
group, i.e. $A,B=1,2,3$, and $f_{AB}(\phi)$ is a symmetric tensor that
is a function of scalar field $\phi$.  It is important to remember that,
in quantum field theories, fluctuations of a field around the vacuum,
i.e. the ground state, are identified as excitations of corresponding
particle.  Accordingly, we should consider an expansion of fields around
the vacuum to treat perturbative quanta.  If the function $f_{AB}(\phi)$
acquires nontrivial vacuum expectation value (VEV) and all of its
eigenvalues are positive-definite no ghosts appear in the gauge
fields. It is plausible to expect that the similar mechanism eliminates
the ghosts in other fields as well.

Now, we see that a Lagrangian of a ghost free theory necessarily
contains nonrenormalizable terms.  However, this should not be a problem
if we consider the theory to be an effective theory of more fundamental
theory, which is perhaps the case for any quantum field theories of
phenomenological interest.  Rather, the presence of nonrenormalizable
terms in the theory merely implies the fact that there is unknown
short-distance structure that is not treated explicitly and there is no
theoretical difficulty \cite{Weinberg:1978kz, Lepage:1997cs}.  For the
reasons that we explained at the beginning of this section and for the
fact that the analysis becomes simple, we consider a construction in the
the framework of $\mathcal{N}=1$ global supersymmetry. 

The structure of the paper is as follows.  In
section~\ref{sec:basicsetup}, we review and introduce materials that are
necessary in the analysis of the models.  In section~\ref{sec:global},
we present two models based on the global $SU(1,1)$, one with two
doublets and the other with a single triplet in order to explain the
problems and the conditions for the superpotential and K\"ahler
potential.  Also, the analysis of the vacuum structure and the
discussion of the transition between degenerate vacua are presented.  In
section~\ref{sec:local}, we consider the gauging of the symmetry and
examine the necessary conditions for the gauge kinetic function. It is
found that, in the case of the local symmetry, two doublet model
considered in this paper is inconsistent. Some remarks about the
obstacles in the construction of the model with noncompact symmetry are
given.  Section~\ref{sec:summary} is devoted to the summary and
discussions of the prospects for the extension of the model.
\section{Basic Setup}
\label{sec:basicsetup}
In this section, we review the basic elements of our framework in order
to elucidate what must be achieved in order to construct a consistent
theory based on a noncompact group. We also briefly review the
transformation properties of the representations of $SU(1,1)$, which are
useful for construction of invariants.

\subsection{$\mathcal{N}=1$ global supersymmetry}
\label{sec:supersymmetry}
The Lagrangian of general $\mathcal{N}=1$ globally supersymmetric theory
with terms up to and including two spacetime derivatives is completely
determined by specifying three functions; the superpotential $W$, the
K\"ahler potential $K$, and the gauge kinetic function $f_{AB}$.  The
gauge kinetic function is introduced only when there is an internal
local symmetry. The superpotential $W(\Phi)$ and the gauge kinetic
function $f_{AB}(\Phi)$ are {\it arbitrary} functions that are
holomorphic in the chiral superfield (there could be more than one),
collectively denoted by $\Phi$, and have mass dimensions three and zero
respectively. The K\"ahler potential $K(\Phi,\Phi^{\dagger})$ is also an
{\it arbitrary} real function of chiral and anti-chiral superfields, and
has mass dimension two.  The mass dimensions here refer to the ones in
four-dimensional spacetime~($D=4$).  Note that both the superpotential
and the K\"ahler potential must be invariant under the symmetry
transformation, while the gauge kinetic function must be constructed so
that it transforms as a symmetric product of two adjoints of the gauge
group.

The kinetic terms for chiral multiplets originate from a K\"ahler
potential, while those for vector multiplets come from a gauge kinetic
function.  The bosonic part of the Lagrangian is given as follows
\begin{eqnarray}
\mathcal{L}_{scalar}
&=&
K^{\bar{i}j}
\left(
D_{\mu}\Phi
\right)^{\dagger}_{\bar{i}}
\left(
D_{\mu}\Phi
\right)_{j}
-V_F
,
\\
V_F
&=&
W^i
\left(
K^{-1}
\right)_{i\bar{j}}
W^{\bar{j}}
,\quad
W^{i}
\equiv
\frac{\partial W}{\partial \Phi_i}
,\quad
W^{\bar{i}}
\equiv
\left(W^{i}\right)^{\dagger},
\label{expandftermpot}
\\
\mathcal{L}_{gauge}
&=&
-\frac{1}{8}
\biggl(
f_{AB}+
\left(
f_{AB}
\right)^{\star}
\biggr)
F_{\mu\nu}^A F^{B \mu\nu}
+\frac{i}{8}
\biggl(
f_{AB}-
\left(
f_{AB}
\right)^{\star}
\biggr)F_{\mu\nu}^A \tilde{F}^{B \mu\nu}
-V_D
,
\\
V_{D}
&=&
\frac{g^2}{4}
\biggl\{
\left(f^{-1}\right)^{AB}
+
\left(f^{-1}\right)^{\star AB}
\biggr\}
\left(
K^i \left(H_A\right)_{i}^{\;\;j}\Phi_j
\right)
\left(
K^{l} \left(H_B\right)_{l}^{\;\;m}\Phi_m
\right),
\quad
K^{i}
\equiv
\frac{\partial K}{\partial \Phi_{i}},
\label{dtermpot}
\end{eqnarray}
where $D_{\mu}$ is a covariant derivative for a corresponding
representation, $g$ is the gauge coupling constant, and $H_{A}$ are
generators in a corresponding representation.  We have also introduced
the K\"ahler metric $K^{\bar{i}j}$; a dimensionless quantity defined as
\begin{eqnarray}
K^{\bar{i}j}
&\equiv&
\frac{\partial^2 K}
{\partial \Phi^{\dagger}_{\bar{i}} \partial \Phi_j}.
\label{kahlermetric}
\end{eqnarray}
The inverses of the K\"ahler metric and gauge kinetic function are
defined by the following
\begin{eqnarray}
K^{\bar{i}j}
\left(K^{-1}\right)_{j\bar{k}}
&=&
\delta^{\bar{i}}_{\;\;\bar{k}},
\quad
f_{AB}\left(f^{-1}\right)^{BC}
=
\delta_{A}^{\;\;B}.
\end{eqnarray}
The indices $i,j,k$ and $\bar{i},\bar{j},\bar{k}$ run over all scalar
fields in the corresponding representations and the indices $A$ and $B$
run over the adjoint representation of the group.  The bars on the
indices represent the conjugates. The raising and lowering of the
indices are to be done using the metric of the symmetry group in the
corresponding representations and the upper and lower indices(with and
without bars) are to be contracted with each other in the standard
manner to form invariants. It should be understood that only the bosonic
components of superfields are to be retained in the formula. The terms
denoted by $V_F$ and $V_D$ are the scalar potentials coming from the
F-term and the D-term respectively.

Let us make a few comments on $V_D$. Using the invariance property of
$W$ and $K$ under the symmetry transformation, we can verify that the
following relations hold
\begin{eqnarray}
\frac{\delta W}{\delta \Phi_{i}}
\left(H_A\right)_{i}^{\;\;j}\Phi_{j}
&=&0
,
\label{relationsuperpot}
\\
\frac{\delta K}{\delta \Phi_{i}}
\left(H_A\right)_{i}^{\;\;j}\Phi_{j}
&=&
\Phi^{\dagger}_{\bar{i}}
\left(H^{\dagger}_A\right)^{\bar{i}}_{\;\;\bar{j}}
\frac{\delta K}{\delta \Phi^{\dagger}_{\bar{j}}},
\label{hermitiankahler}
\end{eqnarray}
Using the relation given in eq.(\ref{hermitiankahler}), we can confirm
that $V_D$ is Hermitian. Note that the potential $V_D$ is present only
in the case of a local symmetry.  It is important to note that, in
general, the K\"ahler metric and the gauge kinetic function are
field-dependent.

As we have explained in section~\ref{sec:intro}, the metrics of the
kinetic terms for the perturbative quanta expanded around the vacuum
must be positive-definite in order for the ghosts to disappear from the
theory.  It is essential to find a K\"ahler potential that gives rise to
a positive-definite K\"ahler metric and a gauge kinetic function, whose
real part is also positive-definite at the vacuum.

As we see from eq.(\ref{dtermpot}) and eq.(\ref{expandftermpot}), the
scalar potentials depend on the real part of $\left(f^{-1}\right)^{AB}$
and $\left(K^{-1}\right)_{i\bar{j}}$ respectively.  This is a special
property of supersymmetric theories; the kinetic term and the potential
term are related to each other through a single function.  This has a
very important consequence.  If the metric of the kinetic term is of an
indefinite one, the potential would be unbounded. The consistency
requires that, if the vacuum exists, the K\"ahler metric evaluated at
the vacuum must be positive-definite.  The analysis of the K\"ahler
metric must be done for presumed VEV's, which are to be determined by
minimizing the potential that itself depends on the K\"ahler metric.
Similar argument applies to the real part of the gauge kinetic function
as well. We should stress that the conditions required for the
positivity of the metric and the ones to ensure the boundedness of the
potential may be different.

Because we consider our theory as an effective theory, we need to
identify the scales in the theory in order to define a low energy
expansion. In this paper, we assume that there are two scales in our
theory; a physical cutoff $M_{high}$, which is typically of order of the
mass of the lightest degrees of freedom that is not treated explicitly
and a low-energy scale $M_{low}$ that is much smaller than $M_{high}$
\footnote{ In principle there could be more than one low-energy mass
scales, but we assume that there is a single low-energy scale for
simplicity.}
. One may imagine that the low energy scale $M_{low}$ is generated
through the dynamics of yet unknown more fundamental theory.  Here we
just assume that such a scale exists in the theory and do not ask its
origin.  We also assume that all parameters in $W(\Phi),
K(\Phi,\Phi^{\dagger})$, and $f_{AB}(\Phi)$ with positive mass dimension
$d>0$ are of $\mathcal{O}(M_{low}^d)$ and the parameters with negative
mass dimension $d<0$ are of $\mathcal{O}(M_{high}^d)$.  The latter
follows from the standard naive dimensional
analysis\cite{Manohar:1983md}.  We should emphasize that the assumptions
here are crucial for our construction.  In particular, the scaling of
couplings with positive mass dimension are essential in order for our
theory to be valid up to the energy scale of the cutoff $M_{high}$.
Therefore it is very important to see whether the assumption are stable
when quantum corrections are taken into account.  However, it is out of
the scope of the present paper and we leave this question to the future
investigation.

As a consequence of supersymmetry, the kinetic terms for chiral fermions
and gaugino are expressed in terms of K\"ahler metric and gauge kinetic
function respectively.  Therefore, if the ghosts in scalar fields and
the gauge fields are eliminated successfully, the ghosts in fermions
disappear automatically.

In order for the field-dependent part of the K\"ahler metric and the
gauge kinetic function to play significant roles, the scalar fields must
acquire nontrivial VEV's
\footnote{In a vacuum that respects Poincare invariance, only scalar
fields are allowed to acquire the VEV's.}
. It means that the consistency requires that the noncompact symmetry
must be spontaneously broken.

To summarize, in order to construct a consistent gauge theory based on a
noncompact local group in a supersymmetric framework, we need to find
the superpotential, K\"ahler potential, and gauge kinetic function that
realize all of the following properties simultaneously: i) The theory
has well-defined vacua. That is, the Hamiltonian is bounded from below.
ii) Only the vacua that break the noncompact symmetry are allowed.  iii)
The K\"ahler metric at the vacuum is positive definite. iv) The real
part of gauge kinetic function at the vacuum is positive-definite.
\subsection{$SU(1,1)$ symmetry}
We consider the transformation properties of the fields in the
fundamental and adjoint representations of $SU(1,1)$, which will be
introduced in our models.

First of all, we define the generators of $SU(1,1)$ to satisfy the
following relations,
\begin{eqnarray}
\left[
H_1,H_2
\right]
=
-iH_3
,\quad
\left[
H_2,H_3
\right]
=
iH_1
,\quad
\left[
H_3,H_1
\right]
=
iH_2.
\label{su11algebra}
\end{eqnarray}
The algebra resembles that of $SU(2)$.  The only difference between the
two is the sign of the right-hand side of the first equation in
eq.(\ref{su11algebra}).

Let us denote a fundamental representation of $SU(1,1)$ by $\Phi$, which
is a doublet.  The transformation law for a doublet is defined as
\begin{eqnarray}
\Phi\to U \Phi,
\end{eqnarray}
where $U$ is an element of $SU(1,1)$ in the fundamental representation
and satisfies
\begin{eqnarray}
U^{\dagger}\eta^{(2)} \;U 
&=&
\eta^{(2)},
\label{su11fund}
\end{eqnarray}
where $\eta^{(2)}$ is the metric of $SU(1,1)$ in the fundamental
representation defined by the following two-by-two matrix,
\begin{eqnarray}
\eta^{(2)}
=
\left(
\begin{array}{cc}
1&0\\
0&-1\\
\end{array}
\right).
\end{eqnarray}
From eq.(\ref{su11fund}), we notice that $U^{-1}=\eta^{(2)}
U^{\dagger}\eta^{(2)}$.  This helps us to see the following
transformation law
\begin{eqnarray}
\Phi^{\dagger}\eta^{(2)}
&\to&
\Phi^{\dagger}\eta^{(2)} U^{-1}.
\end{eqnarray}
The explicit form of $U$ that is useful for examining the transformation
properties is
\begin{eqnarray}
U
&=&
a_0 {\bf 1}_{2\times 2}
+2i \sum_{A=1}^{3}a_A H^{(2)}_A,
\end{eqnarray}
where $a_i\; (i=0,1,2,3)$ are real parameters with a constraint
$a^2_0-a_1^2-a_2^2+a_3^2=1$.  Here, ${\bf 1}_{2\times 2}$ is a
two-by-two unit matrix and $H^{(2)}_A$ are the generators of $SU(1,1)$
in the fundamental representation.  Another representation of $U$ is
given in appendix~\ref{app:paravac}. We choose the following matrices to
represent $H_A$,
\begin{eqnarray}
H^{(2)}_1
&=&
\frac{i}{2}
\left(
\begin{array}{cc}
0&1\\
1&0\\
\end{array}
\right)
,\quad
H^{(2)}_2
=
\frac{1}{2}
\left(
\begin{array}{cc}
0&1\\
-1&0\\
\end{array}
\right)
,\quad
H^{(2)}_3
=
\frac{1}{2}
\left(
\begin{array}{cc}
1&0\\
0&-1\\
\end{array}
\right).
\label{su11generatorfund}
\end{eqnarray}
They satisfy the following normalization condition,
\begin{eqnarray}
{\mbox{tr}}\left(
H^{(2)}_A H^{(2)}_B
\right)
=
\frac{1}{2}\eta^{(3)}_{AB}.
\end{eqnarray}
We can also confirm the relation
$\eta^{(2)}H^{(2)\dagger}_A\eta^{(2)}=H^{(2)}_A$ by using the explicit
expression. Given the explicit form of $U$, we can show the following
transformation law
\begin{eqnarray}
\Phi^T H^{(2)}_2 
\to \Phi^T H^{(2)}_2 U^{-1}.
\end{eqnarray}

Let us denote an adjoint representation of $SU(1,1)$, which is a
triplet, by
\begin{eqnarray}
X&=&
\eta^{AB}_{(3)}X_A H^{(2)}_B 
,
\label{compx}
\end{eqnarray}
where $H^{(2)}_A$ are given by eq.(\ref{su11generatorfund}), and
$\eta^{AB}_{(3)}$ is defined by
\begin{eqnarray}
\eta^{AB}_{(3)}\eta^{(3)}_{BC}
&=&
\delta^{A}_{\;\;C}.
\end{eqnarray}
The triplet $X$ transforms as $X\to U^{-1}X U$, where $U$ is an element
of $SU(1,1)$ defined by eq.(\ref{su11fund}).  We also obtain the
following transformation law by taking the Hermitian conjugate,
\begin{eqnarray}
\left(
\eta^{(2)} X^{\dagger} \eta^{(2)}
\right)
&\to& 
U^{-1}
\left(
\eta^{(2)} X^{\dagger} \eta^{(2)}
\right)U.
\end{eqnarray}
The transformation properties that we presented in this subsection would
be sufficient for constructing invariants and covariant objects from
doublets and triplets.

\section{Models with global $SU(1,1)$ symmetry}
\label{sec:global}
In this section, we present two examples of models with a noncompact
{\it global} symmetry. We show that the models are free from ghosts if
we impose appropriate conditions on the superpotential and K\"ahler
potential.  Note that the gauge kinetic function is not introduced in
the case of a global symmetry.

If we give up the renormalizability in the traditional sense, we are
left with no definite principle for determining the form of the
superpotential, K\"ahler potential or the gauge kinetic function.  In
our construction of the models, we introduce particular type of
functions, which might be unfamiliar.  However, it is probably the
simplest way to realize the properties that are necessary for the
construction without making unreasonable assumptions.  We will give an
argument to support the introduction of inverse type K\"ahler
potentials. Furthermore, it does not seem to pose any immediate problems
as far as we understand.

Before presenting the models, we give a useful formula for a K\"ahler
metric and study, with generality to some extent, the conditions for a
K\"ahler potential that gives rise to a positive-definite K\"ahler
metric. For an illustration of a possibly general property, we consider
a K\"ahler potential that depends on a single triplet of $SU(1,1)$
denoted by $X_{A},\;(A=1,2,3)$ in such a way that its dependence is only
through the variable $x\equiv \eta_{(3)}^{AB}X^{\dagger}_A X_B $, which
is real and invariant under the group. The general form of the K\"ahler
metric for this type of K\"ahler potential can be expressed as
\begin{eqnarray}
K^{AB}(x)
&=&
\frac{\partial^2 K}
{\partial X^{\dagger}_{A} \partial X_B}
=
K'(x)\;\eta_{(3)}^{AB}
+
K''(x)\;
\left(
\eta_{(3)}^{AC}\eta_{(3)}^{BD}
X_{C} X^{\dagger}_{D}
\right),
\label{generalkahlermet}
\end{eqnarray}
where 
\begin{eqnarray}
K'(x)&\equiv&
\frac{d K}{d x}
,\quad
K''(x)\equiv
\frac{d^2 K}{d x^2}.
\end{eqnarray}
It is easy to find that, due to the symmetry property of the group, the
inverse of the K\"ahler metric is expressed as
\begin{eqnarray}
\left(K^{-1}\right)_{AB}
&=&
a(x)\eta^{(3)}_{AB}+b(x)X_{A}X^{\dagger}_{B}
,\quad
a(x)
=
\frac{1}{K'}
,\quad
b(x)
=
-\frac{K''}{K'\left(K'+x K''\right)}.
\label{generalinvkahlermet}
\end{eqnarray}
Now, we use the following property that holds for a general Hermitian
three-by-three matrix to find the conditions to realize a
positive-definite K\"ahler metric. Let $M$ be an Hermitian
three-by-three matrix.  The eigenvalues of $M$ are all real and positive
if and only if it satisfies the following three conditions,
\begin{eqnarray}
\det M >0,\quad
{\mbox{tr}}M >0,\quad
{\mbox{tr}}M^{-1}>0.
\end{eqnarray}
By applying them to the case of the K\"ahler metric given in
eq.(\ref{generalkahlermet}), we obtain the following conditions for a
K\"ahler potential expressed as
\begin{eqnarray}
0&<&\det K^{AB}
=K'^2\left(K'+x K''\right),
\label{detk}
\\
0&<&{\mbox{tr}}K^{AB}
=-K'+K''\left(|X_1|^2+|X_2|^2+|X_3|^2\right),
\label{trk}
\\
0&<&{\mbox{tr}}\left(K^{-1}\right)_{AB}
=-\frac{1}{K'}
-\frac{K''}{K'\left(K'+x K''\right)}
\left(|X_1|^2+|X_2|^2+|X_3|^2\right),
\label{trinvk}
\end{eqnarray}
which then lead to the following conditions
\begin{eqnarray}
0&<&K'+x K'',
\label{condkdet}
\\
\frac{K'+x K''}{2K''}&<&|X_3|^2,\quad
0 < K'',
\label{condktrk}
\\
-\frac{K'}{2K''}&<&|X_3|^2,\quad
K'<0,
\label{condktrkinv}
\end{eqnarray}
correspondingly. The boundary in terms of $x$ where at least one of the
eigenvalues changes the sign is given by the condition $\det K^{AB}=0,\;
\pm \infty$. We will elaborate on the physical implications of the
boundaries later.  It is very important to note that these conditions
are consistent with $SU(1,1)$ symmetry.  This follows from two important
properties; $SU(1,1)$ invariance and the Hermiticity of the matrix.
Therefore, the positivity of the K\"ahler metric is a gauge-invariant
concept.

Let us consider a K\"ahler potential that is a monomial of the form
\begin{eqnarray}
K(x)&=& c x^n,
\label{kahlermono}
\end{eqnarray}
where $c$ is a real constant and $n$ is an integer. It is easy to
confirm that the conditions given in eq.(\ref{condkdet}),
(\ref{condktrk}), and(\ref{condktrkinv}) can be satisfied, namely the
eigenvalues can become all positive, only for $x>0$ and when the
parameters are chosen to be $0<c$ and $n<0$.  This means that the
K\"ahler potential must be of the inverse type.  We can not claim that
the statement applies to a more general class of K\"ahler potentials
from the argument that we presented here, but the similar argument may
lead to such conclusion.  At least, it applies to the case of the
K\"ahler potential that will be introduced in the next subsection.
\subsection{two doublet model}
\label{sec:twodoubletglobal} 
In this subsection, we present a model with the {\it global} $SU(1,1)$
symmetry.  We introduce two chiral superfields $\Psi$ and $\Psi'$ that
transform as the $SU(1,1)$ doublets. The superpotential and the K\"ahler
potential are given by
\begin{eqnarray}
W\left(
\Psi, \Psi'
\right)
&=&
M\phi
+\frac{\lambda }{\phi},
\\
K\left(\Psi,\Psi^{\dagger}, \Psi',\Psi'^{\dagger}\right)
&=&
\frac{g_n}{\left(\Psi^{\dagger}\cdot \Psi\right)^n}
+\frac{g'_n}{\left(\Psi'^{\dagger}\cdot \Psi'\right)^n},
\label{twodoubletglobalspkp}
\end{eqnarray}
where $n$ is an integer to specify a model.  The invariants are given as
follows,
\begin{eqnarray}
\phi
&\equiv&
2\left(
\Psi^T H_2 \Psi'
\right),
\quad
\Psi^{\dagger}\cdot \Psi
\equiv
\Psi^{\dagger}\eta^{(2)}\Psi,
\quad
\Psi^{'\dagger}\cdot \Psi'
\equiv
\Psi^{'\dagger}\eta^{(2)}\Psi'.
\label{twodoubletsiglet}
\end{eqnarray}
Note that we need at least two doublets to write down a superpotential
because of the symmetry property. The dimensional parameters are assumed
to take the values expected from the assumption that we explained in
subsection~\ref{sec:supersymmetry};
\begin{eqnarray}
M&=&
\tilde{M}M_{low}
,\quad
\lambda
=
\tilde{\lambda}M^{5}_{low},
\quad
g_{n}
=
\tilde{g}_{n} M^{2n+2}_{low}
,\quad
g'_{n}
=
\tilde{g}'_{n} M^{2n+2}_{low},
\end{eqnarray}
where dimensionless parameters $\tilde{M},\tilde{\lambda},\tilde{g}_{n}$
and $\tilde{g}'_{n}$ are $O(1)$.

The scalar potential of the model has a contribution only from the
F-term and can be obtained using the formula given in
eq.(\ref{expandftermpot}).  In order to calculate $V_F$, it is useful to
introduce the components of the doublets in the following way
\begin{eqnarray}
\Psi
\equiv
\left(
\begin{array}{c}
\psi_{1}\\
\psi_{2}\\
\end{array}
\right)
,\quad
\Psi'
\equiv
\left(
\begin{array}{c}
\psi'_{1}\\
\psi'_{2}\\
\end{array}
\right).
\label{doubletcomponents}
\end{eqnarray}
The invariants are expressed in terms of these variables as
\begin{eqnarray}
\phi
&=&
\psi_{1}\psi'_{2}-\psi_{2}\psi'_{1},
\label{twodoubletphi}
\\
\Psi^{\dagger}\cdot \Psi
&=&
|\psi_{1}|^2-|\psi_{2}|^2,
\\
\Psi^{'\dagger}\cdot \Psi'
&=&
|\psi'_{1}|^2-|\psi'_{2}|^2.
\end{eqnarray}
The K\"ahler metric for the model is expressed in terms of the
components as
\begin{eqnarray}
K^{\bar{i}j}
&=&
\left(
\begin{array}{cc}
K_{(1)}&0\\
0&K_{(2)}\\
\end{array}
\right),
\nonumber\\
K_{(1)}
&=&
g_n n \left(
\Psi^{\dagger}\cdot\Psi
\right)^{-n-2}
\left(
\begin{array}{cc}
n|\psi_{1}|^2+|\psi_{2}|^2\quad&-(n+1)\psi_{1}\psi^{\star}_{2}\\
-(n+1)\psi^{\star}_{1}\psi_{2}\quad &|\psi_{1}|^2+n|\psi_{2}|^2\\
\end{array}
\right),
\nonumber\\
K_{(2)}
&=&
g'_n n \left(
\Psi'^{\dagger}\cdot\Psi'
\right)^{-n-2}
\left(
\begin{array}{cc}
n|\psi'_{1}|^2+|\psi'_{2}|^2\quad&-(n+1)\psi'_{1}\psi'^{\star}_{2}\\
-(n+1)\psi'^{\star}_{1}\psi'_{2}\quad &|\psi'_{1}|^2+n|\psi'_{2}|^2\\
\end{array}
\right),
\label{twodoubkahlermet}
\end{eqnarray}
where we have assigned
$\Phi_{i}=(\psi_{1},\psi_{2},\psi'_{1},\psi'_{2})$ to be used in the
formula.

In order to determine the vacuum and the conditions for its existence,
we use a simplified parametrization of the fields in our analysis.
Before proceeding, we should make a few comments about this point. The
choice of parametrization can have an important consequence, aside from
the conventional matter, in the case of a noncompact group.  It can be
shown that, by exploiting $SU(1,1)$ symmetry, any configuration of
$\Psi$ can be brought to the form
\begin{eqnarray}
\Psi
=
\left(
\begin{array}{c}
u\\
0\\
\end{array}
\right),
\quad
\Psi'
=
\left(
\begin{array}{c}
v_1 e^{i\beta_1}\\
v_2 e^{i\beta_2}\\
\end{array}
\right),
\label{paramglobaltwodoub}
\end{eqnarray}
if $\Psi$ satisfies the following condition
\begin{eqnarray}
\Psi^{\dagger}\cdot \Psi>0,
\label{condpsi}
\end{eqnarray}
where $u,v_1,v_2,\beta_1$, and $\beta_2$ are real. This condition
implies $u\neq 0$. It is important to note that the condition given here
is imposed in an $SU(1,1)$ invariant manner, and thus it is compatible
with the symmetry. We should also point out that the configurations that
do not obey eq.(\ref{condpsi}) are not related to the ones given in
eq.(\ref{paramglobaltwodoub}) by any choice of parameters of $SU(1,1)$
transformations. It may be seen that the space of configurations in the
case of noncompact group is divided into distinct sectors.  See
appendix~\ref{app:paravac} for details. Because there is no symmetry
transformation left to simplify the parametrization any further, it is
sufficient to deal with only five variables. If they acquire nonzero
VEV's, it means that the symmetry is spontaneously broken. Throughout
the analysis of two doublet models, we assume the condition for the
VEV's given in eq.(\ref{condpsi}).

In the following, we study the K\"ahler metric and the scalar potential
$V_F$ to look for the consistent solutions and determine the conditions
to realize what we need.  First, we substitute the parametrization for
the presumed vacuum given in eq.(\ref{paramglobaltwodoub}) into the
eq.(\ref{twodoubkahlermet}) and examine the eigenvalues.  Assuming that
the VEV's take nonzero values, we obtain the eigenvalues
\begin{eqnarray}
&&
g_n n 
u^{-2(n+1)}
,\quad
g_n n^2 
u^{-2(n+1)}
,
\nonumber\\
\nonumber\\
&&
\frac{1}{2}g'_n n
\left(
v_1^2-v_2^2
\right)^{-(n+2)}
\biggl[
(n+1)(v_1^2+v_2^2)
\pm
\sqrt{
(n+1)^2
\left(v_1^2+v_2^2\right)^2
-4n
\left(v_1^2-v_2^2\right)^2}
\biggr].
\label{doubleteigenv}
\end{eqnarray}
We notice the factor $n$ and $n^2$ along with the rest of the common
factors of the first two eigenvalues listed in eq.(\ref{doubleteigenv}).
It implies that $n$ needs to be positive to realize the
positive-definite K\"ahler metric.  For an arbitrary integer $n$, the
term in the square root is positive for generic values of $v_1$ and
$v_2$.  Thus, all the eigenvalues are guaranteed to be real.  It can be
shown that the terms in the square bracket of the third (with $+$ sign)
and the fourth (with $-$ sign) eigenvalues in eq.(\ref{doubleteigenv})
are positive for positive integer $n$. Furthermore, when $n$ is a
positive even integer and the coupling constants satisfy
\begin{eqnarray}
g_n>0
,\quad
g'_n>0,
\label{positivitycond1}
\end{eqnarray}
all the eigenvalues become positive as long as the VEV's are
nonzero. Next, we consider the F-term contribution to the scalar
potential $V_F$ and examine the VEV's.

By substituting the parametrization given in
eq.(\ref{paramglobaltwodoub}) and using the formula given in
eq.(\ref{expandftermpot}), we obtain the expression for the scalar
potential $V_F\left(u,v_1,v_2,\beta_1,\beta_2\right)$,
\begin{eqnarray}
V_F
&=&
M^2
\biggl\{
\left(K^{-1}\right)_{1\bar{1}} v_2^2
+
\left(K^{-1}\right)_{2\bar{2}} v_1^2
+
\left(K^{-1}\right)_{4\bar{4}} u^2
\biggr\}
\biggl|
1-\frac{\Omega}{\left(uv_2 e^{i\beta_{2}}\right)^2}
\biggr|^2
\nonumber\\
&=&
\frac{M^2 u^2\left(n v_1^2+v_2^2\right)}{g_n g'_n n^2}
\biggl(
g'_n u^{2n}
+
g_n\left(v_1^2-v_2^2\right)^n
\biggr)
\biggl|
1-\frac{\Omega}{\left(uv_2 e^{i\beta_{2}}\right)^2}
\biggr|^2,
\quad
\Omega \equiv \left(\frac{\lambda }{M}\right),
\end{eqnarray}
where the quantity $(uv_2 e^{i\beta_{2}})$ is the VEV of the $SU(1,1)$
invariant $\phi$ defined in eq.(\ref{twodoubletphi}). The explicit
expression for $V_F$ above shows that, when $n$ is a positive even
integer and the coupling constants satisfy eq.(\ref{positivitycond1}),
it is bounded from below.  Also, the eigenvalues of the K\"ahler metric
given in eq.(\ref{doubleteigenv}) are all positive automatically under
these conditions.

The scalar potential of the present model has a remarkable feature, i.e.
$V_F \ge 0$. This is a consequence of supersymmetry. It is interesting
to note that this property is shared even with theories based on
noncompact groups.  However, supersymmetry alone is not sufficient to
ensure the boundedness of the energy.  In fact, the potential could
become unbounded, for instance, when $n$ is a positive odd integer. We
conclude that the property $V_F\ge 0$ shows up only when the K\"ahler
metric is positive-definite.

The next task is to find the vacuum and see whether the VEV's of the
fields take nonzero values as we have assumed. We seek for the vacuum
configurations that give $V_F=0$. First of all, from the explicit
expression for $V_F$, we see that the vacuum is realized only for $u
v_2\neq 0$.  The configurations of the vacua are obtained by solving the
following equation
\begin{eqnarray}
\left|
1-\frac{\Omega}{\left(uv_2 e^{i\beta_{2}}\right)^2}
\right|=0.
\label{twodoubletvac}
\end{eqnarray}
We find the solutions to the equation for both $\Omega>0$ and $\Omega<0$
respectively as
\begin{eqnarray}
uv_2 e^{i\beta_{2}}
&=&
\left|\Omega\right|^{\frac{1}{2}}
e^{i\pi l_{+}},
\quad\quad\; l_{+}=0,1
\quad\quad
\left(\Omega>0\right),
\\
uv_2 e^{i\beta_{2}}
&=&
\left|\Omega\right|^{\frac{1}{2}}
e^{i\frac{\pi}{2}+i\pi l_{-}},
\quad l_{-}=0,1
\quad\quad\left(\Omega<0\right),
\end{eqnarray}
which lead to the relations for the VEV's as follows
\begin{eqnarray}
uv_2&=&\pm \left|\Omega\right|^{\frac{1}{2}}
,\quad
\sin\beta_{2}=0
\quad\quad\quad\quad\left(\Omega>0\right),
\\
uv_2&=&\pm \left|\Omega\right|^{\frac{1}{2}}
,\quad
\cos\beta_{2}=0
\quad\quad\quad\quad\left(\Omega<0\right).
\end{eqnarray}
In either case, only the combination $uv_2$ is fixed by the stationary
condition.  Note that we have $uv_{2}=O(M^{2}_{low})$.  The VEV's $v_1$
and $\beta_{1}$ are allowed to take arbitrary values while giving
$V_F=0$.  Thus, we have degenerate vacua or moduli parametrized by these
variables for fixed values of the parameters of the model.  Note that
each point in the space of degenerate vacua corresponds to the
inequivalent physics.

Now, we have shown that the following is achieved for the present model
by choosing $n$ to be positive even integer and the coupling constants
$g_n$ and $g'_n$ to satisfy eq.(\ref{positivitycond1}): (i) Stable and
degenerate vacua exist. (ii) All the vacua break $SU(1,1)$ symmetry by
having nonzero VEV's of the fields. (iii) The K\"ahler metric at the
vacuum is positive-definite. This guarantees that no ghost appears in
the fermion fields as well. Note that these three properties are
realized simultaneously in a consistent manner. As a result, the model
is free from ghosts as we claimed at the beginning of the section.  We
remark that there is no unbroken symmetry left at the vacuum of the two
doublet model.  We also note that the vacua do not break supersymmetry.

\subsection{one triplet model}
\label{sec:onetripletglobal}
In this subsection, we present another model that has the {\it global}
$SU(1,1)$ symmetry.  We introduce a single chiral superfield $X$ that
transforms as an adjoint representation of the group, i.e. an $SU(1,1)$
triplet. The superpotential and K\"ahler potential of the model are
given by
\begin{eqnarray}
W\left(X\right)
&=&
M\left(X\cdot X\right)
+\frac{\lambda_{m}}{
\left(X\cdot X\right)^{m}},
\label{onetripletw}
\\
K\left(
X,X^{\dagger}
\right)
&=&
\frac{g_n}{\left(X^{\dagger}\cdot X\right)^n},
\label{onetripletk}
\end{eqnarray}
where $m$ and $n$ are integers to specify a model.  The dimensional
parameters $M, \lambda_{m}$, and $g_n$ are assumed to take the values
that follow from the assumption explained in
section~\ref{sec:supersymmetry}. The invariants are given as follows,
\begin{eqnarray}
\left(X\cdot X\right)
&\equiv&
2{\mbox{tr}}\left(X^2\right)
=
\eta_{(3)}^{AB}X_{A}X_{B},
\\
\left(
X^{\dagger}\cdot X
\right)
&\equiv&
2{\mbox{tr}}\left(
\eta^{(2)}X^{\dagger}\eta^{(2)}X
\right)
=
\eta_{(3)}^{AB}X^{\dagger}_{A}X_{B},
\end{eqnarray}
where the components of $X$ are defined in eq.(\ref{compx}).

In the analysis of one triplet model, we make the choice of the
parametrization much the same way as we did in the two doublet model. It
can be shown that, by exploiting the $SU(1,1)$ symmetry, any
configuration of $X$ can be brought to the following form
\begin{eqnarray}
X_A
&=&
\left(
0,v_2,v_3 e^{i\alpha}
\right),
\label{paramglobalonetrip}
\end{eqnarray}
where $v_2,v_3$ and $\alpha$ are real, if it satisfies a certain
condition (see appendix~\ref{app:paravac}).  In order to state the
condition clearly, we express $X_A$ in terms of real and imaginary parts
as $X_A=Y_A+iZ_A$, where $Y_A$ and $Z_A$ are real.  The condition on the
parametrization of the field is expressed as
\begin{eqnarray}
\left(Z\cdot Z\right)
>0,
\end{eqnarray}
which implies 
\begin{eqnarray}
&&
v_{3}\neq 0,\quad
\sin\alpha \neq 0.
\label{restricth}
\end{eqnarray}
No condition is imposed on the real part of $X_A$.  Throughout the
following analysis, we assume the condition given in
eq.(\ref{restricth}).

The K\"ahler metric for the model is obtained as 
\begin{eqnarray}
K^{AB}
&=&
n g_n 
\left(X^{\dagger}\cdot X\right)^{-n-2}
\nonumber\\
&&\times
\left(
\begin{array}{ccc}
n|X_1|^2-|X_2|^2+|X_3|^2&(n+1)X_1 X^{\star}_2&-(n+1)X_1 X^{\star}_{3}\\
(n+1)X^{\star}_1 X_2&-|X_1|^2+n|X_2|^2+|X_3|^2&-(n+1)X_2 X^{\star}_{3}\\
-(n+1)X^{\star}_1 X_3&-(n+1)X^{\star}_2 X_3&|X_1|^2+|X_2|^2+n|X_3|^2\\
\end{array}
\right),
\nonumber\\
\end{eqnarray}
where we have assigned the components as $\Phi_{i}=(X_1,X_2,X_3)$ to be
used in the formula given in eq.(\ref{kahlermetric}).  In order to see
whether the K\"ahler metric at the vacuum can be positive-definite, we
substitute the parametrization given in
eq.(\ref{paramglobalonetrip}). Assuming that $v_2,v_3$, and $\alpha$
take nonzero VEV's, the eigenvalues of the K\"ahler metric evaluated at
these presumed VEV's are found to be
\begin{eqnarray}
&&
n g_n\left(
-v_2^2+v_3^2
\right)^{-n-1},
\nonumber\\
&&\frac{n g_n}{2}
\left(
-v_2^2+v_3^2
\right)^{-n-2}
\nonumber\\
&&\times
\left[
(n+1)
\left(
v_2^2+v_3^2
\right)
\pm
\sqrt{
(n+1)^2
\left(
v_2^2+v_3^2
\right)^2
-4n 
\left(
-v_2^2+v_3^2
\right)^2
}
\right].
\label{tripleteigenv}
\end{eqnarray}
The term in the square root is positive for generic values of $v_2,v_3$
and $\alpha$ for an arbitrary integer $n$, and thus the eigenvalues are
real.  It can also be shown that the terms in the square bracket of the
second($+$ sign) and third($-$ sign) eigenvalues are positive for a
positive integer $n$. However, any choice of the parameters of the model
does not guarantee the positive-definiteness of the K\"ahler metric for
generic $v_2$ and $v_3$. From the general argument given at the
beginning of this section, we know that the parameter of the model must
be chosen to be $g_n>0$ and we need the VEV to satisfy
$(-v_2^2+v_3^2)>0$. For this purpose, we need to examine the actual
VEV's to see if such a situation is realized.

Following the formula given in eq.(\ref{expandftermpot}) and
substituting the parametrization given in eq.(\ref{paramglobalonetrip}),
we obtain the expression for the scalar potential
$V_F\left(v_2,v_3,\alpha \right)$.  In doing this, the formula for the
inverse of the K\"ahler metric given at the beginning of this section
might be useful. The expression for the potential is given as 
\begin{eqnarray}
V_F
&=&
\frac{4M^2}{n^2g_n}
\left(
-v_2^2+v_3^2
\right)^{n+2}
\bigg|
1-\frac{\Omega_{m}}{(-v_2^2+v_3^2e^{i2\alpha})^{m+1}}
\bigg|^2
,\quad
\Omega_{m}
\equiv 
\left(
\frac{m \lambda_{m}}{M}
\right).
\label{onetripletftermpot}
\end{eqnarray}
It is easy to see that, when $n$ is a positive even integer and the
coupling constant satisfies
\begin{eqnarray}
g_n>0,
\label{positivitycondtriplet}
\end{eqnarray}
we have $V_F\ge 0$. Again, this is a consequence of supersymmetry.
Unfortunately, in contrast to the case of the two doublet model, the
conditions that lead to the bounded potential do not guarantee the
positive-definiteness of the K\"ahler metric for generic values of $v_2$
and $v_3$. From the explicit expression for the eigenvalues of the
K\"ahler metric, we see that the VEV of the field must satisfy the
relation
\begin{eqnarray}
\left(X^{\dagger}\cdot X\right)\Big|_{VEV}
&=&
-v_2^2+v_3^2>0,
\label{onetripletposcond}
\end{eqnarray}
in order for the metric to be positive-definite.  Note that the quantity
is the VEV of the $SU(1,1)$ invariant. We need to examine the actual
VEV's to find whether the condition is satisfied at the vacuum.

To do this, we look for solutions to the following equation,
\begin{eqnarray}
\bigg|
1-\frac{\Omega_{m}}{(-v_2^2+v_3^2e^{i2\alpha})^{m+1}}
\bigg|=0.
\end{eqnarray}
The solution for $\Omega_{m}>0$ and $\Omega_{m}<0$ are given as
\begin{eqnarray}
\left(
-v^2_2+v_3^2 e^{i2 \alpha}
\right)
&=&
\left|\Omega_{m}\right|^{\frac{1}{m+1}}
e^{i\frac{2\pi}{m+1}l_{+}}
,\quad\quad\;\;
l_{+}=0,1,\cdots ,m
\quad\quad
\left(\Omega_{m}>0\right),
\label{tripletpo}
\\
\left(
-v^2_2+v_3^2 e^{i2 \alpha}
\right)
&=&
\left|\Omega_{m}\right|^{\frac{1}{m+1}}
e^{\frac{i\pi}{m+1}\left(1+2 l_{-}\right)}
,\quad
l_{-}=0,1,\cdots ,m
\quad\quad
\left(\Omega_{m}<0\right),
\label{tripletno}
\end{eqnarray}
where the term on the left-hand side is the VEV of $SU(1,1)$
invariant $\left(X\cdot X\right)$.  Note that, in order for a solution
of the equation to be a configuration that gives $V_F=0$, it must
satisfy
\begin{eqnarray}
\left(X\cdot X\right)\Big|_{VEV}
&=&
(-v_2^2+v_3^2e^{i2\alpha})\neq 0.
\end{eqnarray}
It is important to note that, because of the condition given in
eq.(\ref{restricth}), we need to see if the solutions are acceptable as
vacuum configurations. In the following, we will be interested in the
case of the superpotential with $m=2$.

Firstly, for $\Omega_{2}>0$, we naively have three types of solutions,
which are specified by the labels $l_{+}=0,1,2$ in
eq.(\ref{tripletpo}). However, we find that the one with $l_{+}=0$ is
not acceptable because of the condition given in
eq.(\ref{restricth}). The rest of the solutions, i.e. those with
$l_{+}=1,2$, is eligible for the vacuum and leads to the following
relations for the VEV's
\begin{eqnarray}
v_3^2
&=&
\sqrt{
\left(v_2^2-\frac{1}{2}|
\Omega_{2}|^{\frac{1}{3}}
\right)^2
+\frac{3}{4}|\Omega_{2}|^{\frac{2}{3}}
},
\quad
\sin(2\alpha)
=
\frac{\pm \frac{\sqrt{3}}{2}
|\Omega_{2}|^{\frac{1}{3}}}
{\sqrt{
\left(v_2^2-\frac{1}{2}|
\Omega_{2}|^{\frac{1}{3}}
\right)^2
+\frac{3}{4}|\Omega_{2}|^{\frac{2}{3}}}}
,
\nonumber\\
\cos(2\alpha)
&=&
\frac{v_2^2-\frac{1}{2}|
\Omega_{2}|^{\frac{1}{3}}}
{\sqrt{
\left(v_2^2-\frac{1}{2}
|\Omega_{2}|^{\frac{1}{3}}
\right)^2
+\frac{3}{4}|\Omega_{2}|^{\frac{2}{3}}
}}
\quad\quad\quad\quad\quad\quad\quad
({\mbox{type F1}}),
\label{typef1}
\end{eqnarray}
which express $v_3$ and $\alpha$ in terms of $v_2$, where $v_2$ is
allowed to take arbitrary values.  We refer to this type of vacua as
``type F1''.  The model possesses degenerate vacua parametrized by a
continuous variable $v_2$. Note that we have $v_3=O(M_{low})$ if we
choose $v_2=O(M_{low})$. Let us look at the eigenvalues of K\"ahler
metric at these vacua.  In order to examine whether the condition given
in eq.(\ref{onetripletposcond}) is satisfied, we evaluate the following
quantity
\begin{eqnarray}
v_3^4-v_2^4
&=&
|\Omega_{2}|^{\frac{1}{3}}
\left(
|\Omega_{2}|^{\frac{1}{3}}-v_2^2
\right).
\end{eqnarray}
Then, it is easy to see that the K\"ahler metric is positive-definite in
the subregion of type F1 vacua that is specified by
\begin{eqnarray}
0\le v_2^2<|\Omega_{2}|^{\frac{1}{3}},
\label{f1region}
\end{eqnarray}
and thus the ghost is absent in this region of the vacua. On the other
hand, in the rest of the region of the type F1 vacua, the K\"ahler
metric is not positive-definite and the ghost does appear in the theory
there.

We have found that the type F1 vacua are separated into two distinct
regions by points specified by $v_2^2=|\Omega_{2}|^{\frac{1}{3}}$ in
eq.(\ref{typef1}); one with a ghost and the other without a ghost.  We
call the former a ``ghost'' phase and the latter a ``ghost free'' phase.
Because the potential energy is the same $V_F=0$ at all points within
the type F1 vacua, one might worry about the possibility of a transition
of VEV's between the two phases. If such a transition is allowed, the
theory would be ill-defined. It turns out that this does not likely to
happen.  We will give the argument for this below.

To explain the basic idea of the argument about the transition in our
model, let us consider a scalar field $\phi$ with a Lagrangian of the
form
\begin{eqnarray}
\mathcal{L}
&=&
\frac{1}{2}K(\phi)\left(\partial_{\mu}\phi\right)^2
-V(\phi),\quad
V(\phi)
=K^{-1}(\phi)f(\phi),
\end{eqnarray}
and the corresponding Hamiltonian given as
\begin{eqnarray}
\mathcal{H}
&=&
\frac{1}{2}K(\phi)
\left[
{\dot{\phi}}^2
+
\left(\nabla \phi\right)^2
\right]+V(\phi).
\end{eqnarray}
Note that the kinetic and potential terms are related to each other
through a single function $K(\phi)$ just as in our supersymmetric
models. We assume that the potential has degenerate minima at $V=0$.  We
also assume that the function $K$ do not have zeros in the space of
vacua, which is similar to the case of our one triplet model, but we may
have points or regions characterized by the VEV $v_{crit}$ that gives
$K^{-1}(v_{crit})=0$ within the vacua. Suppose that a transition from
one point in the space of vacua specified by the VEV $v_{i}$ of $\phi$
to another one specified by $v_{f}$ occurs. Then, it must involve a
configuration $\phi_{fi}$ that connects these two VEV's.  Transitions
within the ghost free region occurs with a configuration that gives
finite Hamiltonian density.  If we consider a transition from one point
in the ghost free phase to another one in the ghost phase, the
corresponding configurations necessarily involve the critical point
$v_{crit}$ at which we have $K^{-1}=0$, and thus formally
$\mathcal{H}\to +\infty$.  This suggest that the transitions that cross
critical points do not take place with finite energy fluctuations. As
the VEV's approach the critical value, the kinetic term dominates over
the potential term and thus, it approaches a free theory.

Let us look at the case of $\Omega_{2}<0$.  Naively, we have three types
of solutions; the ones with $l_{-}=0,1,2$ in eq.(\ref{tripletno}).
Actually, all three of them correspond to the vacuum configuration, and
are classified into the following two types of relations for the VEV's.
Those that are derived from $l_{-}=1$ are
\begin{eqnarray}
v_3^2
&=&|\Omega_{2}|^{\frac{1}{3}}-v_2^2
,\quad
\cos(2\alpha)=-1
\quad\quad\quad\quad
({\mbox{type F2}}),
\end{eqnarray}
which are valid in the region of $v_2$ specified by $0\le v_2^2 \le
|\Omega_{2}|^{\frac{1}{3}}$.  We refer to this as ``type F2''
vacua. Note that we have $v_3=O(M_{low})$ in this region.  We see that
in the region specified by
\begin{eqnarray}
0\le v_2^2<\frac{1}{2}|\Omega_{2}|^{\frac{1}{3}}, 
\end{eqnarray}
the positivity condition given in eq.(\ref{onetripletposcond}) is
satisfied. The K\"ahler metric is positive-definite and thus the ghost
is absent there.  In contrast, in the region
$\frac{1}{2}|\Omega_{2}|^{\frac{1}{3}} < v_2^2 \le
|\Omega_{2}|^{\frac{1}{3}}$ the K\"ahler metric is not positive-definite
and thus the ghost appears.
The relations for the VEV's that are derived from $l_{-}=0,2$ given in
eq.(\ref{tripletno}) are
\begin{eqnarray}
v_3^2 
&=&
\sqrt{ \left(v_2^2+\frac{1}{2}| \Omega_{2}|^{\frac{1}{3}}
\right)^2 +\frac{3}{4}|\Omega_{2}|^{\frac{2}{3}} }, \quad
\sin(2\alpha) 
=
\frac{\pm \frac{\sqrt{3}}{2}
|\Omega_{2}|^{\frac{1}{3}}} {\sqrt{ \left(v_2^2+\frac{1}{2}|
\Omega_{2}|^{\frac{1}{3}} \right)^2
+\frac{3}{4}|\Omega_{2}|^{\frac{2}{3}}}} ,\quad 
\nonumber\\
\cos(2\alpha)
&=&
\frac{v_2^2+\frac{1}{2}| \Omega_{2}|^{\frac{1}{3}}} {\sqrt{
\left(v_2^2+\frac{1}{2} |\Omega_{2}|^{\frac{1}{3}} \right)^2
+\frac{3}{4}|\Omega_{2}|^{\frac{2}{3}} }}
\quad\quad\quad\quad\quad\quad\quad
({\mbox{type F3}}),
\end{eqnarray}
which are also expressed in terms of $v_2$. Note that they are valid for
arbitrary values of $v_2$.  We refer to this as ``type F3'' vacua. For
the type F3 VEV's, the following property is satisfied for an arbitrary
value of $v_2$
\begin{eqnarray}
v_3^4-v_2^4
&=&
|\Omega_{2}|^{\frac{1}{3}}
\left(
|\Omega_{2}|^{\frac{1}{3}}+v_2^2
\right)>0.
\end{eqnarray}
Consequently, the ghosts are absent at every point in the space of type
F3 vacua. We see that $v_3=O(M_{low})$ if we choose $v_2=O(M_{low})$.
For the same reason that we explained about the transition between
regions with and without ghosts, a theory with $\Omega_{2}<0$ in the
ghost free phase do not cross the critical region and thus it is
well-defined.

Our analysis here has demonstrated that the following is achieved by
choosing $n$ to be positive even integer and the coupling constant $g_n$
to satisfy eq.(\ref{positivitycondtriplet}): (i) Stable and degenerate
vacua exist for both $\Omega_2>0$ and $\Omega_{2}<0$ cases. The VEV
$v_2$ is not fixed by the requirement of the minimum energy
condition. We should point out that the vacuum configuration specified
by $v_2=0$ is a part of vacua of the present model.  (ii) All the vacua
break $SU(1,1)$ symmetry. For the vacuum with $v_2\neq 0$, $SU(1,1)$ is
completely broken and no unbroken symmetry is left.  At the vacua with
$v_2=0$, the $U(1)$ symmetry, which is the maximal compact subgroup of
$SU(1,1)$, is left unbroken. (iii) The K\"ahler metric is
positive-definite in the certain subregion of the vacua, which we refer
to as ghost free phase.  The positivity of the K\"ahler metric
guarantees that no ghost appears in the fermion fields.  The transition
between the ghost free phase and the ghost phase are highly suppressed.
These three properties are realized simultaneously in a consistent
manner. As a result, the one triplet model with global $SU(1,1)$
symmetry can be defined without the appearance of ghosts just as we
claimed at the beginning of the section. We also note that supersymmetry
is unbroken at the vacuum.

Although we have not presented the results for the case of
superpotential with $m=1$ in this paper, one can confirm that stable
vacua and the positive-definite K\"ahler metric are obtained.  However,
when the symmetry is made local, no solution exists that realizes
$V_F=0$, $V_D=0$ and the positive-definite metric simultaneously.  This
does not immediately lead to the conclusion that theory is ill-defined,
because the possibility of a vacuum with $V\neq 0$ is not excluded.
However, the analyses become more involved due to the lack of manifest
vacuum, i.e. $V=0$.  Because our purpose of the present paper is to
present an example of a viable model, we consider the one that allows a
simple analysis.
\section{Models with local $SU(1,1)$ symmetry}
\label{sec:local}
In this section, we consider the gauging of the $SU(1,1)$.  In the case
of a local symmetry, it is the sum of the two $V=V_F+V_D$ that we need
to look at in order to examine the stability of the system.  Since we
already have the expression for $V_F$ in each model, what we need to
calculate is $V_D$.  We show that one triplet model become free of
ghosts when certain conditions on the parameters of the superpotential,
K\"ahler potential, and the gauge kinetic function are satisfied.

Before proceeding, we give a useful formula for the inverse of the gauge
kinetic function $f_{AB}$ to be used in the calculation of $V_D$.
Suppose that $f_{AB}$ is a function of fields, collectively denoted by
$\Phi$ of the following form
\begin{eqnarray}
f_{AB}(\Phi)
&=&
-\eta^{(3)}_{AB}+w(\Phi) N_{A}N_{B},
\end{eqnarray}
where $N_{A}$ is a function of $\Phi$ that transforms as an adjoint
representation of the group and $w(\Phi)$ is an invariant function of
$\Phi$. Note that $N_{A}$ is an arbitrary function of $\Phi$, which is
not restricted to be quadratic in $\Phi$.  Then, its inverse
$\left(f^{-1}\right)^{AB}$ is given as
\begin{eqnarray}
\left(f^{-1}\right)^{AB}(\Phi)
&=&
-\eta^{AB}_{(3)}
-\frac{ w}{1-w\left(N\cdot N\right)}
N^{A}N^{B}.
\label{formulainversefab}
\end{eqnarray}

Unfortunately, it seems difficult to state the conditions that ensure
the positivity of the real part of $f_{AB}$, unlike the case of the
K\"ahler metric due to the non-Hermiticity of $f_{AB}$.
\subsection{two doublet model}
\label{sec:twodoubletlocal}
In this subsection, we consider the two doublet model, which we have
studied in subsection~\ref{sec:twodoubletglobal}, but with the local
$SU(1,1)$ symmetry.  The model is described by the same superpotential
and the K\"ahler potential given in eq.(\ref{twodoubletglobalspkp}).
The gauge kinetic function for the model is given by
\begin{eqnarray}
f_{AB}\left(
\Psi, \Psi'
\right)
&=&
-\eta^{(3)}_{AB}
+\xi 
\frac{\left(\Psi \Psi'\right)_A \left(\Psi \Psi'\right)_B}
{\left(\Psi \Psi \Psi' \Psi'\right)_{singlet}},
\end{eqnarray}
where $\xi$ is a dimensionless parameter of the model and the indices
$A$ and $B$ run over the adjoint representation of $SU(1,1)$. For
simplicity, we confine our investigation to the case of real $\xi$. The
covariant and invariant objects are introduced as
\begin{eqnarray}
\left(\Psi \Psi'\right)_A
&\equiv&
4\left(
\Psi^T H^{(2)}_2 H^{(2)}_A \Psi'
\right),
\nonumber\\
&=&
\left(
i\left(
\psi_{1}\psi'_{1}-\psi_{2}\psi'_{2}
\right)
,\;
-\left(
\psi_{1}\psi'_{1}+\psi_{2}\psi'_{2}
\right)
,\;
-\left(
\psi_{1}\psi'_{2}+\psi_{2}\psi'_{1}
\right)
\right)
\label{doulbetadjoint}
\\
\left(\Psi \Psi \Psi' \Psi'\right)_{singlet}
&=&\phi^2,
\end{eqnarray}
where $\phi$ is defined in eq.(\ref{twodoubletsiglet}) and the
components of $\Psi$ and $\Psi'$ are introduced as in
eq.(\ref{doubletcomponents}).  One might notice that the reason for
introducing two doublets in the present model is to form an invariant
that is quartic in fields and to keep $\xi$ a dimensionless parameter.

Before examining whether the real part of gauge kinetic function
evaluated at the vacuum is positive-definite, we substitute the
parametrization given in eq.(\ref{paramglobaltwodoub}) and study their
properties for generic values of $v_1,v_2,\beta_1$ and $\beta_2$.  We
obtain for $f_{AB}(v_1,v_2,\beta_1,\beta_2)$
\begin{eqnarray}
f_{AB}
&=&
\left(
\begin{array}{ccc}
1&0&0\\
0&1&0\\
0&0&-1\\
\end{array}
\right)
+\frac{\xi}{v_2^2}
\left(
\begin{array}{ccc}
-v^2_1 e^{i2(\beta_1-\beta_2)}&
-iv^2_1 e^{i2(\beta_1-\beta_2)}&
-iv_1 v_2 e^{i(\beta_1-\beta_2)}\\
-iv_1^2 e^{i2(\beta_1-\beta_2)}&
v^2_1 e^{i2(\beta_1-\beta_2)}&
v_1 v_2 e^{i(\beta_1-\beta_2)}\\
-iv_1 v_2 e^{i(\beta_1-\beta_2)}&
v_1 v_2 e^{i(\beta_1-\beta_2)}&
v^2_2 \\
\end{array}
\right).
\label{twodoubletfab}
\end{eqnarray}
The eigenvalues of the real part of $f_{AB}$ are given as follows
\begin{eqnarray}
1-\xi \left(\frac{v_1}{v_2}\right)^2
,\quad
\frac{1}{2v_2^2}
\biggl[
(v_1^2+v_2^2)\xi
\pm
\sqrt{(\xi-2)^2v_2^4
+\xi^2 v_1^4
+2v_1^2v_2^2\xi(\xi+2)
}
\biggr],
\end{eqnarray}
for $2(\beta_{1}-\beta_{2})=l\pi$, where $l$ is an integer. It can be
shown that, for generic values of $v_1$ and $v_2$, any choice of $\xi$
does not guarantee the positivity of the eigenvalues. Note that we give
the explicit form of the eigenvalues only for
$2(\beta_{1}-\beta_{2})=l\pi$, but the results are basically the same
for other values.  We need to find the minimum of $V=V_F+V_D$ and
examine the VEV's of the fields to see whether the positive-definite
metric can be realized.

By substituting the parametrization given in
eq.(\ref{paramglobaltwodoub}) into the formula given in
eq.(\ref{dtermpot}) and after somewhat lengthy calculations, we obtain
for $V_D(u,v_1,v_2,\beta_1,\beta_2)$
\begin{eqnarray}
V_D
&=&
\frac{g^2}{8r}
\biggl[
\biggl\{
2(-rv_1^2+v_2^2)
\Big(
1+ng'_n\left[v_1^2-v_2^2\right]^{-(n+1)}
\Big)
+
u^2
\left(1+ng_n u^{-2(n+1)}\right)
\biggr\}^2
\nonumber\\
&&-4rv_1^2
\left(
v_1^2 \xi \cos(2\beta)-v_2^2
\right)
\Big(
1+ng'_n\left[v_1^2-v_2^2\right]^{-(n+1)}
\Big)^2
\biggr],
\label{twodoubdterm}
\end{eqnarray}
where $r\equiv \xi-1,\;\beta\equiv \beta_{1}-\beta_{2}$, with the help
of the formula for the inverse of $f_{AB}$ given at the beginning of
this section.

The term on the second line of eq.(\ref{twodoubdterm}) does not have a
definite sign irrespective of the choice of $n$ and of the coupling
constants of the model $g_n,g'_n$ and $\xi$. It is a manifestation of a
general property of theories based on a noncompact group that there
appear directions in the field space that decrease the potential energy
arbitrarily.  In order for the potential to be bounded from below, those
directions must disappear somehow. However, because such direction can
not be eliminated by any choice of the parameters for the present case,
the D-term potential $V_D$ is not bounded from below.  This provides
another example that illustrates the fact that supersymmetry alone does
not guarantee the boundedness of the potential.  The two doublet model
with local $SU(1,1)$ symmetry does not possess a ground state and hence
it can not be defined consistently.
\subsection{one triplet model}
\label{sec:onetripletlocal}
In this subsection, we consider the model with one triplet, which we
studied in subsection~\ref{sec:onetripletglobal}, but with the local
$SU(1,1)$ symmetry.  The superpotential and K\"ahler potential of the
model are given by eqs.(\ref{onetripletw}) and (\ref{onetripletk}). The
gauge kinetic function for the model is
\begin{eqnarray}
f_{AB}\left( X \right) &=& -\eta^{(3)}_{AB} +\xi \frac{X_A X_B}
{\left(X\cdot X\right)},
\label{tripletgaugekin}
\end{eqnarray}
where $\xi$ is a dimensionless parameter and the indices $A$ and $B$ run
over the adjoint representation of $SU(1,1)$. For simplicity, we confine
ourself to the case of real $\xi$.

Just as in the case of the two doublet model, we first study the
properties of the gauge kinetic function for generic VEV's of the fields
and examine the positivity.  To do this, we substitute the
parametrization given in eq.(\ref{paramglobalonetrip}) into
eq.(\ref{tripletgaugekin}) and obtain for $f_{AB}(v_1,v_2,v_3,\alpha)$
\begin{eqnarray}
f_{AB} &=& \left(
\begin{array}{ccc}
1&0&0\\
0&1&0\\
0&0&-1\\
\end{array}
\right)
+R
\left(
\begin{array}{ccc}
0&0&0\\
0&v_2^2 &v_2 v_3 e^{i\alpha}\\
0&v_2 v_3e^{i\alpha}&v_3^2 e^{i2\alpha}\\
\end{array}
\right),
\quad R\equiv
\frac{\xi}{\left(-v_2^2+v_3^2\right)}.
\label{onetripletfab}
\end{eqnarray}
The eigenvalues of the real part of $f_{AB}$ are expressed as
\begin{eqnarray}
&&1,\quad
\frac{1}{2}
\biggl[
R
\left(
v_2^2+v_3^2\cos(2\alpha)
\right)
\pm
\sqrt{
\left(
2-Rv_3^2 \cos(2\alpha)
\right)^2
+R^2v_2^2 
\left(
v_2^2+2v_3^2
\right)
+4Rv_2^2
}
\biggr].
\label{onetripletrealfab}
\end{eqnarray}
Again, it is not possible to guarantee the positivity of the eigenvalues
for generic values of $v_2,v_3$ and $\alpha$ by any choice of $\xi$.  We
need to look into the actual VEV's to find out whether it can be
realized.

Let us find the minimum of the total potential $V=V_F+V_D$ and examine
the VEV's of the fields. We already have the expression for $V_F$ and
the configurations that are determined by $V_F=0$ alone.  What we need
to do is to obtain the explicit expression for $V_D$.  It is carried out
by substituting the parametrization given in
eq.(\ref{paramglobalonetrip}) into the formula given in
eq.(\ref{dtermpot}), and after somewhat lengthy calculations, we obtain
the expression for $V_D(v_2,v_3,\alpha)$ as
\begin{eqnarray}
V_D&=&
\frac{2n^2 g^2 g_n^2}{(-v_2^2+v_3^2)^{2(n+1)}}
v_2^2v_3^2\sin^2\alpha,
\label{onetripletdtermpot}
\end{eqnarray}
where we used the following expressions for the generators of $SU(1,1)$
in the adjoint representation $H^{(3)}_{A}$, 
\begin{eqnarray}
H^{(3)}_1
&=&
\left(
\begin{array}{ccc}
0&0&0\\
0&0&+i\\
0&+i&0\\
\end{array}
\right)
,\quad
H^{(3)}_2
=
\left(
\begin{array}{ccc}
0&0&-i\\
0&0&0\\
-i&0&0\\
\end{array}
\right)
,\quad
H^{(3)}_3
=
\left(
\begin{array}{ccc}
0&-i&0\\
+i&0&0\\
0&0&0\\
\end{array}
\right).
\label{su11adjoint}
\end{eqnarray}
Note that $V_D$ does not depend on $\xi$.  We immediately see that, when
$n$ is a positive integer and $g_n>0$, $V_D$ is bounded from below for
generic values of $v_2,v_3$ and $\alpha$. Therefore, the conditions that
give rise to the bounded $V_F$ (see the conditions just below
eq.(\ref{onetripletftermpot})) also ensure that $V_D$ is bounded from below. To be
more precise, we have $V_D\ge 0$, which implies $V\ge 0$.  Thus, the
minimum of the total potential could be realized as $V_F=V_D=0$, which
leads to the supersymmetric vacuum.  In the following analysis, we
choose the parameters of the model that lead to bounded $V_F$, i.e. $n$
is a positive even integer and $g_n>0$.

Taking into account the condition given in eq.(\ref{restricth}), we find
that there is only one type of configurations that realizes the minimum
of $V_D$, that is $v_2=0$.  This means that the subgroup $U(1)$ is left
unbroken at the vacuum. In contrast to the requirement $V_F=0$, the
VEV's $v_3$ and $\alpha$ are not fixed by $V_D=0$ and are allowed to be
arbitrary. In the following, we present the analysis of the VEV's for
both $\Omega_{2}>0$ and $\Omega_{2}<0$ cases combining the conditions
from $V_F=0$ and $V_D=0$ including the evaluation of the eigenvalues of
$f_{AB}$ at each vacuum.

For the superpotential with $\Omega_2>0$, there is only one type of
vacuum configurations.  It is given by type F1 that we examined in
subsection~\ref{sec:onetripletglobal} with $v_2=0$.  We see that the
space of vacua parametrized by $v_2$ in the case of the global symmetry
shrinks to points specified by $v_2=0$ and certain discrete values of
$\alpha$ in the case of the local symmetry. We refer to this as ``type
D1'' vacua, which is specified by the following relations
\begin{eqnarray}
v_2=0,\quad
v^2_3=|\Omega_{2}|^{\frac{1}{3}},\quad
\cos(2\alpha)=-\frac{1}{2}
\quad\quad\quad\quad
({\mbox{type D1}}).
\end{eqnarray}
We have already shown that the type D1 vacua give rise to the
positive-definite K\"ahler metric (see eq.(\ref{f1region})).
The eigenvalues of $f_{AB}$ given in eq.(\ref{onetripletrealfab}) at
these vacua are given as
$\left\{1,\;1,\;\frac{1}{2}\left(-2-\xi\right)\right\}$.  Obviously,
with the choice of the parameter $\xi<-2$, all the eigenvalues become
positive at all points in the type D1 vacua and thus the ghost is absent
from the theory.

For the superpotential with $\Omega_2<0$, there are two types of
degenerate vacua. One is of type F2 with the condition $v_2=0$, which is
specified by the following
\begin{eqnarray}
v_2=0,\quad
v^2_3=|\Omega_{2}|^{\frac{1}{3}},\quad
\cos(2\alpha)=-1
\quad\quad\quad\quad
({\mbox{type D2}}).
\end{eqnarray}
We refer to this as ``type D2'' vacua. The eigenvalues of $f_{AB}$ at
these vacua are $\left\{1,\;1,\;\left(-1-\xi\right)\right\}$, which
become all positive for the choice of the parameter $\xi<-1$.
The other is of type F3 with the condition $v_2=0$ and expressed as
\begin{eqnarray}
v_2=0,\quad
v^2_3=|\Omega_{2}|^{\frac{1}{3}},\quad
\cos(2\alpha)=\frac{1}{2}
\quad\quad\quad\quad
({\mbox{type D3}}).
\end{eqnarray}
We refer to these as ``type D3''. The eigenvalues of $f_{AB}$ at these
vacua are $\left\{1,\;1,\;\frac{1}{2}\left(\xi-2\right)\right\}$, which
become all positive with the choice of the parameter $2 <\xi$. For the
case $\Omega_{2}<0$, any choice of the parameter $\xi$ can not realize
the situation that there is no ghost phase.  There are two phases in the
space of vacua.  The transition is expected to be highly suppressed for
the same reason that we explained in
subsection~\ref{sec:onetripletglobal}, and therefore the theory in the
ghost free phase remains so.

We have shown that metrics for the kinetic terms for all the
perturbative quanta are positive-definite if the appropriate parameters
are chosen for the model.  As another important point, we need to make
sure that there are no problems with the mass terms for the gauge
bosons.  We find that the mass terms for the gauge bosons are the same
for all types of vacua and they are given as
\begin{eqnarray}
\mathcal{L}^{gauge}_{mass}
&=&
ng^2 g_n 
\left|\Omega_{2}\right|^{-\frac{n}{3}}
\left[
\left(A^{1}_{\mu}\right)^2
+
\left(A^{2}_{\mu}\right)^2
\right].
\end{eqnarray}
Clearly, they have the correct signs at the vacua, which are necessary
for consistency of the theory. We see that the masses of the gauge
bosons are $O(M_{low})$. The absence of the mass term for $A^{3}_{\mu}$,
which corresponds to gauge field of the compact part of the group,
indicates that the gauge boson remains massless as expected from the
property of the vacua of the one triplet model.

We would like to make a comment about the form of D-term contribution to
the scalar potential. In particular, we point out that an important
element that leads to bounded $V_D$ is the availability of the symmetry
transformations that take generic parametrization of the field into the
one given in eq.(\ref{paramglobalonetrip}).  To see this, let us
consider the most general parametrization of the field
\begin{eqnarray}
X_{A}
&=&
\left(
v_1 e^{i\alpha_{1}}
,
v_2 e^{i\alpha_{2}}
,
v_3 e^{i\alpha_{3}}
\right),
\end{eqnarray}
where $v_i$ and $\alpha_{i}, \;(i=1,2,3)$ are real.  The expression for
$V_D(v_1,v_2,v_3,\alpha_{1},\alpha_{2},\alpha_{3})$ is given by
\begin{eqnarray}
V_D
&=&
\frac{2n^2 g^2 g_n^2 }
{\left(
-v_1^2-v_2^2+v_3^2
\right)^{2(n+1)}}
\nonumber\\
&&\times
\biggl[
v_1^2 v_3^2 \sin^2(\alpha_1-\alpha_3)
+
v_2^2 v_3^2 \sin^2(\alpha_2-\alpha_3)
-
v_1^2 v_2^2 \sin^2(\alpha_1-\alpha_2)
\biggr].
\end{eqnarray}
Apparently, the third term has a negative sign, which might lead to the
unbounded potential.  However, due to the existence of a symmetry
transformation that allows us to set $\alpha_{1}=\alpha_{2}$, it
actually does not lead to instability.  Of course, we can reach this
conclusion by careful inspection of the explicit expression for
$V_D$. Note that the symmetry transformation that we mention here is a
part of those that allow the parametrization given in
eq.(\ref{paramglobalonetrip}).

Our analysis here has demonstrated that the following is achieved for
the model by choosing $n$ to be positive even integer and the coupling
constant $g_n$ to satisfy eq.(\ref{positivitycondtriplet}): (i) Stable
and degenerate vacua exist.  (ii) All the vacua break $SU(1,1)$
symmetry. However, the $U(1)$ symmetry, which is a maximal subgroup of
$SU(1,1)$ is left unbroken.  (iii) The K\"ahler metric at each vacuum is
positive-definite.  (iv) The real part of gauge kinetic function at each
vacuum is positive-definite.  Because of these properties, no ghosts
appear in the fermionic sector as well as in the bosonic sector.  The
transition from the ghost free phase to the ghost phase is not expected
to occur. These three properties are realized simultaneously in a
consistent manner. As a result, the one triplet models(a class of models
specified by an integer $n$) are expected to be free from ghosts as we
claimed at the beginning of the section. We also note that supersymmetry
is not broken at each vacuum.

\section{Summary and discussions} 
\label{sec:summary}
We have presented a construction of theories with a linearly realized
$SU(1,1)$ symmetry, which is the simplest noncompact nonabelian group,
in the framework of $\mathcal{N}=1$ global supersymmetry. In our
construction, the symmetry is realized linearly without introducing the
constrained fields. We first explained the problem of ghosts and
discussed what must be achieved in order to solve it.  For illustration
of important points, we have presented two models both in global and
local symmetry cases.  In the global symmetry case, we have shown that
both the two doublet and one triplet models satisfy all the requirements
for the consistency; The symmetry is spontaneously broken,
positive-definite K\"ahler metric is realized, and the energy is bounded
from below.  In the local symmetry case, two doublet model is shown to
be inconsistent due to the lack of bounded potential.  On the other
hand, we have shown that the one triplet model satisfies all the
requirements for consistency.  Thus our analysis provides a suggestive
evidence that it is possible to define a gauge theory based on a
noncompact group. However, there is an important point that we must
emphasize, i.e. the assumptions on the dimensional coupling constants,
which we explained in section~\ref{sec:supersymmetry}, are the crucial
elements for our construction.  Hence the results of our analysis
heavily depends on them.  If these assumptions are shown to be
inconsistent, our theory would be able to describe only the massless
particles because all the massive particles would have masses of order
of the cutoff $M_{high}$.  Even in that case, our construction is still
useful for description of the dynamics of the massless particles while
realizing the symmetry linearly on the fields.

Although we have constructed models in the framework of $\mathcal{N}=1$
supersymmetry, the supersymmetry may not be a necessity for constructing
a ghost free theory with noncompact group.  Rather, the incorporation of
supersymmetry itself makes it difficult due to the severe restriction it
imposes.  For example, the potential term and the kinetic term are
related to each other through a single function. However, this property
plays an extremely important role in preventing the potential disaster
from occurring.  Due to the existence of ghost phase and ghost free
phase, which have the same potential energy, there is a possibility of
transition between them. In the supersymmetric framework, however, if we
manage to construct a consistent model at the classical level, the phase
transition is highly suppressed because it must involve a configuration
that requires an infinite amount of energy. It is due to the linkage
between the kinetic term and the potential term that the transitions are
suppressed.

Perhaps, we should mention a few words about the inverse type potentials
that we introduced in our construction.  Because of their peculiar form,
which has not been studied in the literature, one might even suspect
that such potentials are not allowed in quantum field theories,
e.g. they might violate locality or causality.  At present, we do not
have any theoretical argument to show that this is not the case.
However, it is interesting to know that there are examples of such type
of potentials that are generated by instanton effects in supersymmetric
QCD \cite{Affleck:1983rr,Affleck:1983mk} though in different context.

We should also mention that our motivation for adopting the inverse type
functions for the superpotential and that for the K\"ahler potential and
gauge kinetic function are essentially different. The reason for the
former is as follows. It has been chosen so that no vacua preserving the
noncompact symmetry are allowed in the theory.  To realize this, the
inverse type potential is probably the simplest choice.  However, we
suspect that such condition is not necessary. What is actually needed is
the existence of the symmetry breaking vacuum in the theory.  The
coexistence of the broken and unbroken vacua does not cause any problem
because transitions between them are suppressed.  The reason for the
latter is that this type of function is probably the only choice to
realize the positive-definite metric without making unreasonable
assumptions. If not for this type of functions, we might have to assume
that the terms that would have been suppressed by large mass scales
$M_{high}$ make sizable contributions in order to reverse the sign of
the metric. Note also that we have explicitly shown with certain
generality that the inverse type function is necessary in the case of
one triplet model in section~\ref{sec:global}. The construction of
nonsupersymmetric theories also must involve the inverse type functions
in the kinetic terms as well.

In order to clarify our perspective on our construction of the models,
we give some comments about the terms that are allowed by the symmetry
of the theory but not included in our potentials. A canonical term in
the K\"ahler potential is one of them.  It gives rise to canonical
kinetic terms for the fields.  Another example is the series of inverse
power terms of different powers.  The inclusion of those terms could
shift the VEV's by $O(1)$ factor, but we may expect the theory to remain
the same qualitatively, i.e. the theory remains free of ghosts. From the
analysis presented in this paper, it should not be hard to imagine that
availability of ghost free theory would not be lost just by including
those terms, especially given the fairly loose conditions for the
coupling constants.  The higher dimensional terms are not included as
well. Their effects are expected to be small. For these reasons our
simple example may be sufficient for a demonstration of a possibility
that a consistent theory with noncompact gauge symmetry can be
constructed.

We have discussed the obstacles that we encounter in the construction.
One of the difficulties is related to the instability. The noncompact
nature of the group generally gives rise to directions in the field
space that destabilize the system.  In order to construct a model with
bounded potentials, we need to find a way to eliminate such
directions. For F-term potentials, it seems easy to eliminate them by
choosing the appropriate K\"ahler potentials.  On the other hand,
finding bounded D-term potentials by searching for positive-definite
gauge kinetic function seems difficult.  One way to avoid such
instability is to construct a model so that these directions are related
to the gauge transformations. We conjecture that this is possible when
certain matter contents are chosen.

The analysis presented in this paper are confined to the study of
classical properties, i.e. the leading order of perturbative
expansions. One might wonder whether quantum corrections may change them
drastically and the whole arguments become invalid. However, if the
perturbation theory is well-behaved, we may expect otherwise. What
concerns us the most about quantum corrections is whether our
assumptions are compatible with them.  Since the potentials we employed
are of a quite unfamiliar type (the inverse type) we do not know what to
expect with certainty.  To the best of our knowledge, no literature
exists in which quantum corrections in such theories are studied.
Therefore, in order to see whether our assumptions are consistent, it is
desirable to study the nature of quantum corrections, especially the
renormalization of K\"ahler potentials. Even if the assumption is shown
to be incompatible with quantum corrections, it does not necessarily
mean the breakdown of our construction. It should not be unreasonable to
expect the possibility that the parameters of the theory are
finely-tuned due to the dynamics of more fundamental theory in such a
way that the dimensional coupling constants behave as our assumptions.

\appendix
\section{Parametrization of the vacuum}
\label{app:paravac}

In this appendix, we elaborate on the choice of the parametrization of
the vacuum that we used in eq.(\ref{paramglobaltwodoub}) and
eq.(\ref{paramglobalonetrip}).  

\subsection{doublet}
First, we consider the case of an $SU(1,1)$ doublet $\Psi$. Let us start
with an arbitrary configuration
\begin{eqnarray}
\Psi
&=&
\left(
\begin{array}{c}
\psi_{1}
\\
\psi_{2}
\end{array}
\right),
\end{eqnarray}
where $\psi_{1}$ and $\psi_{2}$ are complex.  We show that there exists
an $SU(1,1)$ transformation $U$ that takes $\Psi$ into the form
\begin{eqnarray}
\Psi'
&=&
U\Psi
=
\left(
\begin{array}{c}
u
\\
0
\end{array}
\right)
\label{parametpsi},
\end{eqnarray}
where $u$ is real. This can be done by using the explicit expression for
$U$ given as 
\begin{eqnarray}
U
&=&
\left(
\begin{array}{cc}
F&G^{\star}\\
G& F^{\star}\\
\end{array}
\right),
\label{su11matrix1}
\end{eqnarray}
where $F$ and $G$ are complex valued parameters with a constraint
$|F|^2-|G|^2=1$.  We look for a solution for $F$ and $G$ to the
following equations,
\begin{eqnarray}
u&=&F\psi_{1}+G^{\star}\psi_{2},
\quad
0=G\psi_{1}+F^{\star}\psi_{2}.
\label{doubletparameq}
\end{eqnarray}
It is easy to verify that the unique solution to
eq.(\ref{doubletparameq}) exists.  It is given as
\begin{eqnarray}
F
&=&
\frac{\psi^{\star}_{1}}{\sqrt{|\psi_{1}|^2-|\psi_{2}|^2}}
,\quad
G
=
\frac{-\psi_{2}}{\sqrt{|\psi_{1}|^2-|\psi_{2}|^2}}.
\end{eqnarray}
It should be emphasized that the solution exists only for configurations
that satisfy $|\psi_{1}|^2-|\psi_{2}|^2>0$. For this transformation, we
have
\begin{eqnarray}
u&=&
\sqrt{|\psi_{1}|^2-|\psi_{2}|^2}.
\end{eqnarray}
We have used up all three of the $SU(1,1)$ transformation parameters.

\subsection{triplet}
Next, we consider the case of an $SU(1,1)$ triplet $X_{A}$ and confirm
our statement that we made about eq.(\ref{paramglobalonetrip}). Let us
start with an arbitrary configuration,
\begin{eqnarray}
X_{A}&=&
\left(
X_1, X_2,X_3
\right),
\end{eqnarray}
where $X_{A}$ are complex.  The transformation law for $X_{A}$ under
$SU(1,1)$ can be written as 
\begin{eqnarray}
X_{A}&\to&D_{A}^{\;\;B}X_B,
\end{eqnarray}
where $D_{A}^{\;\;B}$ is a real matrix that satisfies
\begin{eqnarray}
\eta^{(3)}D^T \eta^{(3)}D=1.
\end{eqnarray}
The explicit form of $D$ is given by exponentiating the generators in
the adjoint representation given in eq.(\ref{su11adjoint}).  For
convenience, we introduce the following matrices each of which
corresponds to a transformation generated by $H_{A}$
\begin{eqnarray}
D_{A}(\phi_{A})
&=&
e^{i\phi_{A}H^{(3)}_A},\quad
A=1,2,3,
\end{eqnarray}
where $\phi_{A}$ is real and the summation over the index $A$ is not
implied. 

Our argument proceeds in three steps. First, we focus on the imaginary
parts of $X_A$, which are introduced in
section~\ref{sec:onetripletglobal} and denoted by $Z_{A}$. Note that the
components $Z_{A}$ are all real.  Because the real and imaginary parts
never mix with each other by the $SU(1,1)$ transformations, we can treat
them separately. We show that a transformation that takes the arbitrary
configuration of $Z_{A}=(v_1,v_2,v_3)$ into the form
$Z_{A}=(0,v'_{2},v'_{3})$ exists, i.e. the solution to the following
equations exists,
\begin{eqnarray}
\left(
\begin{array}{c}
0\\
v'_2\\
v'_3\\
\end{array}
\right)
=
D_{3}(\phi_3)
\left(
\begin{array}{c}
v_1\\
v_2\\
v_3\\
\end{array}
\right),
\label{t3transf}
\end{eqnarray}
where $D_{3}(\phi_{3})$ is given by
\begin{eqnarray}
D_3(\phi_3)
&=&
\left(
\begin{array}{ccc}
a &b&0\\
-b&a&0\\
0&0&1\\
\end{array}
\right),\quad
a=\cos\phi_3,\quad
b=\sin \phi_3.
\end{eqnarray}
Note that $a$ and $b$ satisfy $a^2+b^2=1$. The solution for $a$ and $b$
to the eq.(\ref{t3transf}) is given as
\begin{eqnarray}
a&=&
\frac{v_2}{\sqrt{v_1^2+v_2^2}},\quad
b=\frac{-v_1}{\sqrt{v_1^2+v_2^2}},
\end{eqnarray}
in which case we have
\begin{eqnarray}
v'_2&=&
\sqrt{v_1^2+v_2^2},\quad
v'_3=v_3.
\end{eqnarray}
Note that there are no restrictions on $v_1$ and $v_2$.  Next, we show
that there exists a transformation that takes arbitrary configuration of
$Z_{A}=(0,v'_2,v_3)$ into the form $Z_{A}=(0,0,v''_{3})$, i.e. the
solution to the following equations exists,
\begin{eqnarray}
\left(
\begin{array}{c}
0\\
0\\
v''_3\\
\end{array}
\right)
=
D_{1}(\phi_1)
\left(
\begin{array}{c}
0\\
v'_2\\
v_3\\
\end{array}
\right),
\label{t1transf}
\end{eqnarray}
where $D_{1}(\phi_{1})$ is given by
\begin{eqnarray}
D_1(\phi_1)
&=&
\left(
\begin{array}{ccc}
1 &0&0\\
0&c&d\\
0&d&c\\
\end{array}
\right),\quad
c=\cosh\phi_1,\quad
d=\sinh\phi_1.
\end{eqnarray}
Note that $c$ and $d$ satisfy $c^2-d^2=1$.  It is easy to verify that an
unique solution to eq.(\ref{t1transf}) exists and is given as
\begin{eqnarray}
c
&=&
\frac{v_3}{\sqrt{v_3^2-v'^{2}_{2}}}
,\quad
d=\frac{-v'_2}{\sqrt{v_3^2-v'^{2}_{2}}}
\end{eqnarray}
We stress that the solution exists only for configurations that satisfy
$v_3^2-v'^{2}_{2}>0$, which translates into $-v^2_{1}-v^2_{2}+v^2_{3}>0$
in the original variables. We arrive at $X_{A}$ of the following form
\begin{eqnarray}
X_{A}
&=&
\left(
\begin{array}{c}
u_{1}\\
u_{2}\\
u_{3}\\
\end{array}
\right)
+i
\left(
\begin{array}{c}
0\\
0\\
v''_3\\
\end{array}
\right),
\quad
v''_3=
\sqrt{-v^2_{1}-v^2_{2}+v^2_{3}},
\end{eqnarray}
where $u_1,u_2,$ and $u_3$ are real. We consider further transformation
by $D_3$, which does not bring any changes to the imaginary part $Z_{A}$
(See eq.(\ref{t1transf})). With the transformation $D_{3}(\phi'_{3})$
given as
\begin{eqnarray}
D_3(\phi'_3)
&=&
\left(
\begin{array}{ccc}
a' &b'&0\\
-b'&a'&0\\
0&0&1\\
\end{array}
\right),\quad
a'=
\frac{u_2}{\sqrt{u_1^2+u_2^2}}
,\quad
b'=\frac{-u_1}{\sqrt{u_1^2+u_2^2}},
\end{eqnarray}
we can finally bring it to the form
\begin{eqnarray}
X_{A}
&=&
\left(
\begin{array}{c}
0\\
u'_{2}\\
u_{3}\\
\end{array}
\right)
+i
\left(
\begin{array}{c}
0\\
0\\
v''_3\\
\end{array}
\right),
\quad
u'_{2}
=\sqrt{u^2_{1}+u^2_{2}}.
\end{eqnarray}

\begin{acknowledgments}
The authors would like to thank K.~Harada for helpful discussions. One of
us (N.Y.) would like to thank the organizers and participants of the
16th Yukawa International Seminar(YKIS), especially, M.~Bando, T.~Kugo,
K.~Yamawaki, and R.~Kitano for useful comments.  This work was supported
in part by a Grant-in-Aid for Scientific Research on Priority Areas
($\sharp$ 441) ``Progress in elementary particle physics of the 21st
century through discoveries of Higgs boson and supersymmetry''
(No. 16081209) from the Ministry of Education, Culture, Sports, Science
and Technology of Japan. 

\end{acknowledgments}

\bibliographystyle{apsrev}
\bibliography{refnoncomp.bib}

\end{document}